%% file: main.tex
\documentclass[sigconf, screen]{acmart}

\setcopyright{none}
\renewcommand\footnotetextcopyrightpermission[1]{}

\acmConference[HotCarbon'25]{Workshop on Sustainable Computer Systems}{July 10-11th, 2025}{Cambridge, MA}

\usepackage[normalem]{ulem}
\usepackage{enumitem}
\usepackage{rotating}
\usepackage{hyperref}
\usepackage{booktabs}
\usepackage{balance}
\usepackage{graphicx}
\usepackage{wrapfig}
\usepackage[ruled, vlined, linesnumbered]{algorithm2e}
\usepackage{amsmath}
\usepackage{fancyhdr}
\usepackage{bbm}
\usepackage{multirow}
\usepackage{xcolor}
\usepackage{soul}
\usepackage{tikz}
\usetikzlibrary{arrows.meta, positioning, shapes.geometric}

\usepackage{cleveref}
\usepackage{url}

\usepackage{mathtools}
\usepackage{pifont}
\usepackage{makecell}
\usepackage{subcaption}
\usepackage{xspace}
\usepackage[skins]{tcolorbox}
\usepackage{array}
\usepackage[super]{nth}

\newcommand{\PreserveBackslash}[1]{\let\temp=\\#1\let\\=\temp}
\newcolumntype{C}[1]{>{\PreserveBackslash\centering}p{#1}}
\newcolumntype{R}[1]{>{\PreserveBackslash\raggedleft}p{#1}}
\newcolumntype{L}[1]{>{\PreserveBackslash\raggedright}p{#1}}
\def\BibTeX{{\rm B\kern-.05em{\sc i\kern-.025em b}\kern-.08em
		T\kern-.1667em\lower.7ex\hbox{E}\kern-.125emX}}

\setlist{noitemsep, leftmargin=*, topsep=0pt, partopsep=0pt}

\crefformat{section}{\S#2#1#3} 
\crefformat{subsection}{\S#2#1#3}
\crefformat{subsubsection}{\S#2#1#3}

\tcbset{
	takeaway/.style={ enhanced,width=\hsize,left=0pt,right=0pt,top=0pt,bottom=0pt,colback=black!20!blue!4!white,boxrule=1pt,colframe=black!30!blue!20!white
	},
}
\newcounter{takeawaycounter}
\newcommand{\takeawaybox}[1]{
	\stepcounter{takeawaycounter}
	\begin{tcolorbox}[takeaway]
		\textbf{Takeaway~\arabic{takeawaycounter}:}#1
	\end{tcolorbox}
}

\tcbset{
	conclusion/.style={ enhanced,width=\hsize,left=0pt,right=0pt,top=0pt,bottom=0pt,colback=black!20!green!4!white,boxrule=1pt,colframe=black!30!green!20!white
	},
}
\newcounter{conclusioncounter}

\newcommand{\SYSTEM}{SCARF}

\begin{document}

\title{Not All Water Consumption Is Equal: A Water Stress Weighted Metric for Sustainable Computing}


\author{Yanran Wu}
\orcid{0009-0009-9025-4170}
\affiliation{%
  \institution{Purdue University}
   \city{West Lafayette}
   \state{IN}
   \country{USA}}
\email{wu2187@purdue.edu}

\author{Inez Hua}
\orcid{0000-0003-4977-5758}
\affiliation{%
   \institution{Purdue University}
   \city{West Lafayette}
   \state{IN}
   \country{USA}}
\email{hua@purdue.edu}

\author{Yi Ding}
\orcid{0000-0003-2757-9182}
\affiliation{%
   \institution{Purdue University}
   \city{West Lafayette}
   \state{IN}
   \country{USA}}
\email{yiding@purdue.edu}

\input{tex/0-abstract}

\ccsdesc[500]{Social and professional topics~Sustainability}
\ccsdesc[500]{Hardware~Impact on the environment}
\ccsdesc[500]{Computing methodologies~Machine learning}

\keywords{Sustainability, Water Consumption, Large Language Model, Semiconductor Manufacturing, Datacenter}

\thanks{This work is accepted to present at the 4th Workshop on Sustainable Computer Systems (HotCarbon'25). The proceeding version is published in ACM SIGEnergy Energy Informatics Review (EIR), Volume 5 Issue 2, July 2025.}

\maketitle

\input{tex/1-introduction}

\input{tex/2-method}

\input{tex/3-case1}

\input{tex/4-case2}
\input{tex/5-case3}
\input{tex/6-conclusion}

\bibliographystyle{ACM-Reference-Format}
\bibliography{reference}

\end{document}

%% file: tex/0-abstract.tex
\begin{abstract}

Water consumption is an increasingly critical dimension of computing sustainability, especially as AI workloads rapidly scale. However, current water impact assessment often overlooks where and when water stress is more severe. To fill in this gap, we present \SYSTEM{}, the first general framework that evaluates water impact of computing by factoring in both spatial and temporal variations in water stress. \SYSTEM{} calculates an Adjusted Water Impact (AWI) metric that considers both consumption volume and local water stress over time. Through three case studies on LLM serving, datacenters, and semiconductor fabrication plants, we show the hidden opportunities for reducing water impact by optimizing location and time choices, paving the way for water-sustainable computing. The code is available at \url{https://github.com/jojacola/SCARF}.
\end{abstract}

%% file: tex/1-introduction.tex
\section{Introduction}

The rapid growth of computing demand, particularly in AI workloads, has significantly increased water consumption across the whole computing stack, intensifying water-related sustainability concerns~\cite{mit2025genAIimpact, Ali2024making}.  
Recent studies show that serving 10 to 50 medium-length queries from a large language model (LLM) such as GPT-3 can consume 500 mL of water~\cite{Ali2024making}. Datacenters, which power AI services, consume vast amounts of water for cooling. One Google datacenter in Iowa consumed about 3.7 million cubic meters of water in a year~\cite{google2024sustainability}, and according to Bloomberg, nearly two-thirds of new U.S. datacenters since 2022 are located in high water-stress regions~\cite{nicoletti2025aiwater}. Semiconductor manufacturing is also water-intensive, as large volumes of ultra-pure water are used to clean silicon wafers. 
For instance, Intel~\cite{intelCSR2024} reported its annual global water consumption of 9.6 million cubic meters.
Globally, water consumption by all sectors has increased
exponentially over the past century~\cite{kummu2016world}. This has led to conflicts between stakeholders, including the computing industry. For example, in Taiwan, TSMC has competed with agriculture for water, especially during drought years~\cite{zhong2021drought, nyt2021taiwandrought}.

\emph{Water consumption} refers to the volume of water that is removed from the immediate environment, including water that is evaporated, transpired, or incorporated into products or processes~\cite{macknick2011review, reig2013s}. This metric differs from \emph{water withdrawal}, which includes all water extracted from the sources, including that discharged back after use. Due to this removal from local watershed, \emph{water consumption} is more indicative of assessing regional water impact. 
Following prior work~\cite{Ali2024making, protocol2011greenhouse}, in this paper, we focus on \emph{Scope 1} water consumption (i.e., water consumed directly on-site, such as in cooling systems) and \emph{Scope 2} (i.e., off-site water consumption associated with electricity generation).  \emph{Scope 3} (i.e., water consumed in upstream supply chain) is not discussed due to the lack of public data.

The environmental impact of water consumption depends heavily on spatial and temporal context. 
Spatially, the same amount of water consumption can lead to greater environment burden in arid regions than in water-abundant regions. For instance, both Central California and the Mississippi River Basin are major farming regions, but California's dry climate has caused more rapid groundwater level decline and thousands of dry wells~\cite{jasechko2024rapid, latimes2024groundwater, yasarer2020trends}. Temporally, water supply fluctuates over time. For example, California experiences distinct wet and dry seasons, and climate projections indicate these fluctuations will become more pronounced in the future~\cite{mallakpour2018new}. Therefore, we must account for both the geographic location and the temporal trajectory when analyzing water impact.

Prior work has explored water impact in computing, estimating water consumption of AI workloads such as LLM training and inference~\cite{morrisonholistically, desroches2025exploring, Ali2024making, waterIR}, as well as datacenters~\cite{lei2025water, shehabi20242024, gupta2024dataset, gnibga2024flexcooldc}. These efforts primarily focus on modeling and profiling the amount of water consumed by computational tasks or facility operations, providing valuable insights of water demand across different computing layers. In parallel, others have begun to incorporate regional water conditions into system design, including infrastructure deployments~\cite{siddik2021environmental, chen2022characterizing} and workload scheduling~\cite{islam2015water, islam2016exploiting, li2024towards, waterwise}. These studies aim to reduce water impact by integrating water consumption as a factor in facility siting or computing workload management.

However, existing efforts face two key limitations. First, many focus solely on the quantity of water consumed, ignoring spatial variability in water stress~\cite{morrisonholistically, Ali2024making, waterIR, lei2025water, shehabi20242024, gupta2024dataset, gnibga2024flexcooldc, islam2015water, islam2016exploiting, li2024towards}. Second, studies that incorporate regional water data often rely on static or short-term information~\cite{desroches2025exploring, siddik2021environmental, chen2022characterizing, waterwise}, failing to account for how water stress evolves over time due to climate change or long-term resource depletion. These two limitations ignore where and when water stress occurs, leading to an incomplete picture of water impact. Thus, there is an urgent need for a spatial- and temporal-aware framework to evaluate water impact in computing.




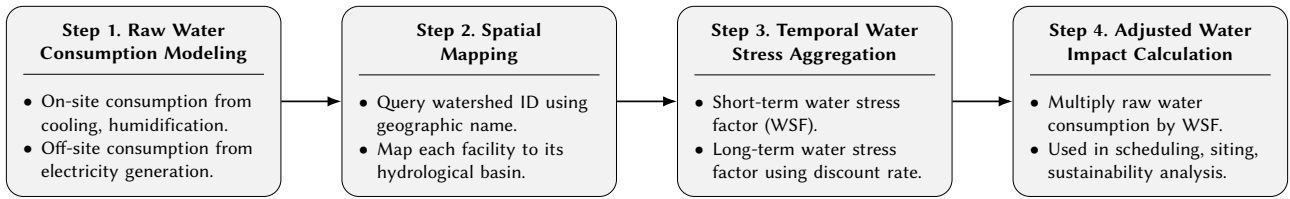
\begin{figure*}[t]
\centering
\scalebox{0.8}{
\begin{tikzpicture}[
    node distance=1.6cm and 1.0cm,
    mybox/.style={
        draw, fill=gray!10, rounded corners=8pt,
        minimum height=3cm,
        minimum width=4.2cm,
        text width=4.0cm, align=left, font=\sffamily,
        inner sep=8pt,
    },
    arrow/.style={-{Latex}, thick, color=black}
]

\node[mybox] (step1) {
    \begin{minipage}[t][2.6cm][t]{\linewidth}
        \centering \textbf{Step 1. Raw Water Consumption Modeling} \\[-0.4em]
        \rule{\linewidth}{0.4pt} \\[0.5em]
        \raggedright
        \begin{itemize}
            \item On-site consumption from cooling, humidification.
            \item Off-site consumption from electricity generation.
        \end{itemize}
    \end{minipage}
};

\node[mybox, right=of step1] (step2) {
    \begin{minipage}[t][2.6cm][t]{\linewidth}
        \centering \textbf{Step 2. Spatial \\ Mapping} \\[-0.4em]
        \rule{\linewidth}{0.4pt} \\[0.5em]
        \raggedright
        \begin{itemize}
        \item Query watershed ID using geographic name. 
        \item Map each facility to its hydrological basin.
        \end{itemize}
    \end{minipage}
};

\node[mybox, right=of step2] (step3) {
    \begin{minipage}[t][2.6cm][t]{\linewidth}
        \centering \textbf{Step 3. Temporal Water Stress Aggregation} \\[-0.4em]
        \rule{\linewidth}{0.4pt} \\[0.5em]
        \raggedright
        \begin{itemize}
        \item Short-term water stress factor (WSF).
        \item Long-term water stress factor using discount rate.
        \end{itemize}
    \end{minipage}
};

\node[mybox, right=of step3] (step4) {
    \begin{minipage}[t][2.6cm][t]{\linewidth}
        \centering \textbf{Step 4. Adjusted Water Impact Calculation} \\[-0.4em]
        \rule{\linewidth}{0.4pt} \\[0.5em]
        \raggedright
        \begin{itemize}
        \item Multiply raw water consumption by WSF.
        \item Used in scheduling, siting, sustainability analysis.
        \end{itemize}
    \end{minipage}
};

\draw[arrow] (step1.east) -- (step2.west);
\draw[arrow] (step2.east) -- (step3.west);
\draw[arrow] (step3.east) -- (step4.west);

\end{tikzpicture}
}
\caption{Overview of the \SYSTEM{} Framework.}
\label{fig:scarf}
\end{figure*}

To bridge this gap, we present \SYSTEM{} (\underline{\textbf{S}}tress-\underline{\textbf{C}}orrected \underline{\textbf{A}}ssessment of Water \underline{\textbf{R}}esource \underline{\textbf{F}}ootprint), a general framework for evaluating water impact from computing. The key insight is to integrate spatial and temporal variation in \emph{water stress}, which is the ratio of local water demand to supply, reflecting regional scarcity. \SYSTEM{} follows four steps: (1) model water consumption volume, including both on-site (Scope 1) and off-site (Scope 2) consumption; (2) map each site to its corresponding hydrological basin; (3) calculate a Water Stress Factor (WSF) using cumulative projections over time; and (4) compute the Adjusted Water Impact (AWI) by multiplying total water consumption with the WSF, yielding a unified metric for water sustainability. To demonstrate the effectiveness and generalizability of \SYSTEM{}, we present three case studies covering LLM serving, datacenters, and semiconductor fabrication plants. Our key takeaways for sustainable computing deployment include:

\begin{itemize}
    \item \emph{LLM serving: } The adjusted water impact of deploying LLMs is highly location-sensitive and can vary significantly due to seasonal water supply changes. 
    \item \emph{Datacenters: } The adjusted water impact of datacenters depends on regional water stress, consumption volume, and weighting of long-term versus short-term water stress.
    \item \emph{Semiconductor fabrication plants: } Fab plants located in high water stress regions can impose a much higher adjusted water impact than those in low-stress regions. 
\end{itemize}
We summarize the contributions as follows:
\begin{itemize}
    \item Introducing \SYSTEM{}, the first systematic and general framework for evaluating the water impact with water stress in computing.
    \item Designing a unified metric that incorporates both spatial and regional variations in water stress, enabling sustainable planning of the computing infrastructure.
    \item Conducting case studies in LLM serving, datacenters, and semiconductor fab plants, enabling comparison for water impact evaluation across the full stack of computing.
\end{itemize}

%% file: tex/2-method.tex
\section{The \SYSTEM{} Framework}

\Cref{fig:scarf} shows the four steps of \SYSTEM{}. The key insight is to assess the environmental impact of water consumption by factoring in where and when water stress occurs. \SYSTEM{} models total water consumption (on-site and off-site), maps site locations to hydrological basins, calculates a Water Stress Factor (WSF) by aggregating water stress over time, and computes the Adjusted Water Impact (AWI) by applying the WSF. We detail each step next.


\subsection{Step 1: Raw Water Consumption Modeling}

We first model the raw operational water consumption, which includes both on-site (Scope 1) and off-site (Scope 2) consumption, following the same methodology from prior work~\cite{Ali2024making, waterIR, morrisonholistically}.

\textbf{On-site water consumption} refers to water that is directly consumed at the facility, including cooling, humidification, and other operational needs. It can be quantified using the \emph{on-site Water Usage Effectiveness} ($\text{WUE}_{\text{on}}$) metric in the unit of liters/kWh:
\begin{equation}
    W_{\text{on}} = P \cdot t \cdot \text{WUE}_{\text{on}},
    \label{eq:w_on}
\end{equation}
where $W_{\text{on}}$ is the on-site water consumption (in liters), $P$ is the average power draw (in kilowatts), and $t$ is the runtime (in hours).

\textbf{Off-site water consumption} refers to the indirect water consumption associated with electricity generation at power plants. This can be modeled by first computing the total energy used and then applying the water intensity of upstream electricity sources, denoted as the \emph{off-site Water Usage Effectiveness} ($\text{WUE}_{\text{off}}$) metric in the unit of liters/kWh:
\begin{equation}
    W_{\text{off}} = P \cdot t \cdot \text{PUE} \cdot \text{WUE}_{\text{off}},
    \label{eq:w_off}
\end{equation}
where $W_{\text{off}}$ is the off-site water consumption (in liters), and $\text{PUE}$ is the \textit{Power Usage Effectiveness} calculated by dividing the total energy used by the facility by the energy used by its IT equipment.




    

\subsection{Step 2: Spatial Mapping}

To account for spatial variation in water stress, we map the physical location of each facility to its corresponding hydrological basin, where water stress is actually measured~\cite{kuzma2023aqueduct}. Since water stress varies by watershed---not by country or region---the same facility in two different watersheds can face different water challenges (see \Cref{fig:ms_datacenters}). To perform this mapping, we first use the facility's geographic name (e.g., city or administrative region) to query the Aqueduct API~\cite{wri_aqueduct_api}, which returns the corresponding watershed ID. These watershed IDs from the global hydrographic classification~\cite{lehner2013global} are unique for each hydrological basin. We then use this watershed ID to retrieve both current and projected water stress values from the Aqueduct 4.0 dataset~\cite{kuzma2023aqueduct}. Aqueduct 4.0 is the latest iteration of World Resource Institute's water risk framework to translate complex hydrological data into water risk indicators, including water stress levels that will be used in the next step.



\subsection{Step 3: Temporal Water Stress Aggregation}


To account for temporal variation in water stress, we introduce the Water Stress Factor (WSF)—a time-weighted aggregation of basin-level water stress values. To reflect future uncertainty and potential impact, WSF applies a user-defined discount rate $\gamma$, following principles from environmental economics~\cite{weitzman1994environmental, tol2011social}. A higher discount rate reduces the weight of future stress; an infinitely large rate ignores future stress entirely. Below, we derive both short-term and long-term WSFs to support different application needs.

\textbf{Short-Term WSF.} For short-term WSF analyses, such as evaluating the immediate water stress of computing tasks like LLM serving, we focus on the current impact only:
\begin{equation}
\label{eq:scf_short}
WSF_b^{\text{short}} = WS_{t_0, b},
\end{equation}
where $WS_{t_0,b}$ denotes the water stress at basin $b$ at time $t_0$.

\textbf{Long-Term WSF.} For long-term facilities like datacenters, which operate for decades, factoring in future water stress provides a more accurate view of sustainability, as regional water availability may change over time. However, future conditions are uncertain due to factors like climate variability and technological shifts. To account for this and weigh future impacts appropriately, we apply a discounting approach inspired by environmental economics~\cite{weitzman1994environmental, tol2011social}:
\begin{equation}
\label{eq:scf_long}
WSF_b^{\text{long}} = \sum_{t \in T} w_t \cdot WS_{t,b},
\end{equation}
where $WS_{t,b}$ denotes the water stress at basin $b$ at time $t$. The weight $w_t$ is computed using a discount rate $\gamma$, reflecting the relative importance of future water stress:
\begin{equation}
w_t' = \frac{1}{(1 + \gamma)^{t - t_0}}, \quad w_t = \frac{w_t'}{\sum_{t' \in T} w_{t'}'}
\end{equation}
where $t_0$ is the baseline year, and $T$ is the set of years considered. According to Aqueduct 4.0~\cite{kuzma2023aqueduct}, the baseline year is 2019, representing current conditions based on over 40 years of hydrological data (1979–2019), and $T=\{\rm baseline, 2030, 2050, 2080\}$. A lower discount rate gives more weight to future water stress, while a higher rate prioritizes near-term impacts. Note that when $\gamma \to \infty$ where near-short impacts are maximized, all \( w_t' \) tend to zero except for \( t = t_0 \), where \( w_{t_0}' = 1 \). In this case, the equation will be reduced to the short-term WSF in \Cref{eq:scf_short}.





\subsection{Step 4: Adjusted Water Impact Calculation}

After aggregating water stress at the basin level, we compute the \textbf{Adjusted Water Impact} (AWI) by multiplying the total raw water consumption by the corresponding Water Stress Factor (WSF). AWI captures both onsite and offsite water consumption associated with water stress levels in basin $b$. Depending on the time horizon, we apply either the short-term or long-term WSF.
\begin{align}
\label{eq:awi}
\text{AWI} = (W_{\text{on}} + W_{\text{off}}) \times WSF_{b}
\end{align}

%% file: tex/3-case1.tex
\section{Case Study I: LLM Serving}\label{sec:case1}


In this section, we use \SYSTEM{} to evaluate the \underline{short-term} water impact of LLM serving. LLM serving is flexible and dynamic; cloud providers can shift regions, scale deployments, or update models within months~\cite{qwen2,qwen2.5}. Because of this short lifecycle, the adjusted water impact depends mainly on the immediate time and location of execution, not long-term trends.

\subsection{Evaluation Methodology}


We run three Qwen2.5 models (7B, 14B, 32B)~\cite{qwen2.5} on a server with an NVIDIA H100 GPU and Intel Xeon 8480+ CPU. We assume the workload runs in Microsoft datacenters across different regions, where on-site WUEs and PUEs are publicly reported~\cite{microsoft_datacenter_efficiency}. Location-specific off-site WUEs are obtained from the Water Impact Tool~\cite{siddik2024waterintensity}.

\textbf{Power measurement.} We collect real-time power consumption data $P$ for the GPU and CPU during inference. Power readings are sampled every 200ms using NVIDIA's \verb|pynvml| API for the GPU and Intel's \verb|psutil| API for the CPU.

\textbf{Workload and profiling.} We conduct experiments on the open-source LLM serving platform vLLM~\cite{kwon2023efficient} and simulate user queries with requests sampled from the ShareGPT dataset~\cite{sharegpt}, which contains real multi-turn conversations. For each model, we gradually increase the query-per-second (QPS) rate from 1 to 30. We identify the QPS that delivers peak throughput, then record the power $P$ and latency $t$ for each request at that QPS.

\textbf{Adjusted water impact.} Once we have power $P$, execution time $t$, PUE, and both on-site and off-site WUEs, \SYSTEM{} calculates the raw water consumption based on \Cref{eq:w_on} and \Cref{eq:w_off}. Then, \SYSTEM{} computes the final adjusted water impacts using \Cref{eq:awi} for each model size at different locations.

\subsection{Evaluation Results}

\begin{figure}[!t]
    \centering
    \includegraphics[width=\linewidth]{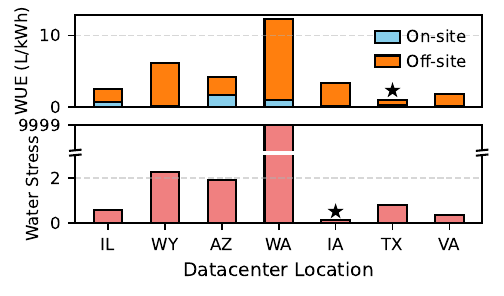} 
    \caption{On-site and Off-site WUEs of Microsoft datacenter locations (top) and their corresponding local water stress values (bottom). \ding{72} refers to the location with lowest value.}
    \label{fig:ms_datacenters}
\end{figure}

\begin{figure}[!t]
    \centering
    \includegraphics[width=\linewidth]{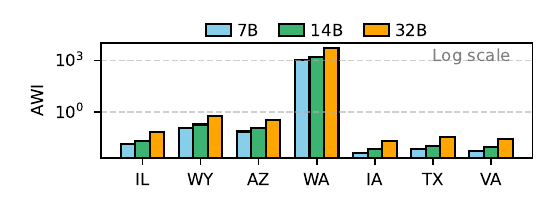}
    \caption{Adjusted Water Impact (AWI) per request for different model sizes in different Microsoft datacenter locations.}
    \label{fig:ms_models_wis}
\end{figure}

\subsubsection{Adjusted water impact in different locations.}

We first compare the adjusted water impact across different locations. In \Cref{fig:ms_datacenters}, the top chart shows on-site and off-site WUEs for datacenters by location. The Texas (TX) datacenter has the lowest WUE, meaning it consumes the least water per unit of electricity. The bottom chart shows water stress levels, with Iowa (IA) having the lowest water stress. These charts reveal large differences in both operational efficiency and environmental conditions across locations.

We then compute adjusted water impact per request by model size using \SYSTEM{}. As shown in \Cref{fig:ms_models_wis}, the same model can have drastically different adjusted water impacts depending on where it is deployed. For example, the datacenter in Quincy, Washington (WA)---a high-desert area---faces high water stress and less efficient WUE, leading to over 1000$\times$ higher adjusted water impact per request than in low-stress locations. Comparing Illinois (IL) and Wyoming (WY), serving a small 7B model in WY can have a higher adjusted water impact than running a large 32B model in IL.

\takeawaybox{
The adjusted water impact of deploying LLMs is highly location-sensitive. Same workloads can have orders-of-magnitude differences in adjusted water impact depending on where they are served. 
}

\subsubsection{Adjusted water impact in different months}

\begin{figure}[!t]
    \centering
    \includegraphics[width=\linewidth]{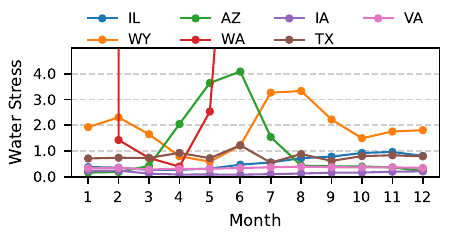}
    \caption{Water stress levels across various Microsoft datacenter locations over the 12 months of the baseline year 2019.}
    \label{fig:ms_monthly_water_stress}
\end{figure}

\begin{figure}[!t]
    \centering
    \includegraphics[width=\linewidth]{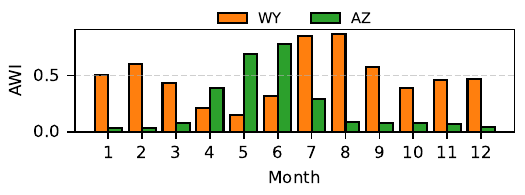}
    \caption{Adjusted water impact (AWI) per request of 32B model in two locations: Wyoming (WY) and Arizona (AZ).}
    \label{fig:ms_32b}
\end{figure}

Water stress at the same location can change month to month due to seasonal shifts in climate and water availability. As shown in \Cref{fig:ms_monthly_water_stress}, water stress fluctuates over time, with locations like Wyoming (WY) and Arizona (AZ) showing particularly large swings.

We therefore select WY and AZ to examine how adjusted water impact changes over time for the same model. \Cref{fig:ms_32b} shows the monthly adjusted water impact per request for a 32B model. From January to March, AZ has relatively lower water stress, resulting in lower adjusted water impact than WY. But during the arid months of April to June, AZ's water stress spikes. As a result, its adjusted impact surpasses WY's, even reaching more than twice as much.

\takeawaybox{
Even in the same location, seasonal changes can significantly affect adjusted water impact. When you deploy matters---not just where.
}

%% file: tex/4-case2.tex
\section{Case Study II: Datacenters}\label{sec:case2}


In this section, we use \SYSTEM{} to evaluate the \underline{long-term} water impact of datacenters. Since datacenters operate for decades~\cite{donovan2013fundamentals, equinix2013dc11}, their adjusted water impact must account for both current stress and future projections.

\subsection{Evaluation Methodology}\label{sec:datacenter_method}
 
We analyze the annual water consumption of Google's U.S. datacenters. We focus on Google because it provides the most detailed public reporting among major cloud providers~\cite{google2024sustainability, microsoft2024sustainability, amazon2023aws}, including site-specific data on annual water consumption and PUE~\cite{google2024sustainability}.

\textbf{Energy consumption estimation.} Since Google does not disclose total energy consumption per datacenter, we estimate each site's power capacity using public data from the industry database \texttt{DataCenters.com}~\cite{datacenterscom}. For each datacenter, we identify all nearby sites within a 100-mile radius and take the maximum reported power capacity $P$ (kWh) as a proxy for the Google site's power capacity. We note that datacenters do not continuously operate at peak power. According to Google's datacenter power dataset~\cite{sakalkar2020data}, the average power utilization of datacenters is approximately 70\%. Assuming 24/7 operation, we compute the annual energy consumption as $E = P \times 24 \times 365 \times 0.7$. We then apply the reported Power Usage Effectiveness (PUE) to estimate the total energy consumption.

\textbf{Adjusted water impact.} Given the decades-long lifespan of datacenters, \SYSTEM{} evaluates the annual adjusted water impact using both current water stress data and future projections for 2030, 2050, and 2080. Aqueduct 4.0 provides three pathways to represent future scenarios of climate and socioeconomic change~\cite{kuzma2023aqueduct}: \emph{Business-as-Usual} pathway with moderate emissions and adaptation, \emph{Optimistic} scenario with sustainable development, and \emph{Pessimistic} scenario under high emissions. We adopt business-as-usual pathway in this analysis. We apply a discount rate of 3\% in long-term water stress factor calculation, following the U.S. Water Resources Development Act of 1974 (WRDA)~\cite{wrda1974}. We also analyze how changes in discount rates and future scenarios affect long-term adjusted water impact, helping to understand sustainability trade-offs under uncertainty.




\subsection{Evaluation Results}

\begin{figure}[!t]
    \centering
    \includegraphics[width=\linewidth]{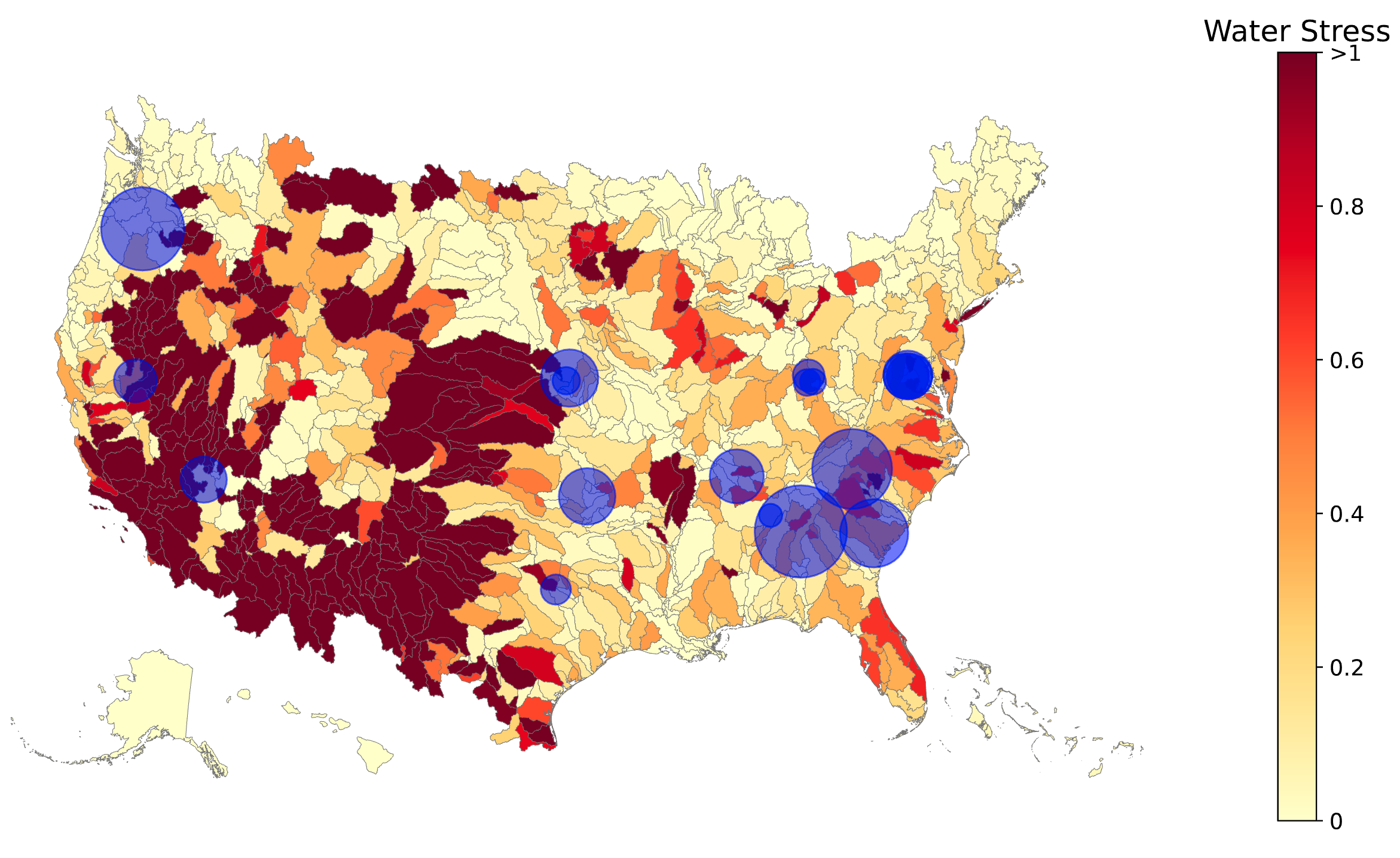}
    \caption{Geographic distribution of water stress (background) and raw water consumption of Google's U.S. datacenters (blue circles). Circle size indicates annual water consumption volume. }
    \label{fig:us_watershed_stress_with_alaska_and_google_circles}
\end{figure}

\begin{figure}[!t]
    \centering
    \includegraphics[width=\linewidth]{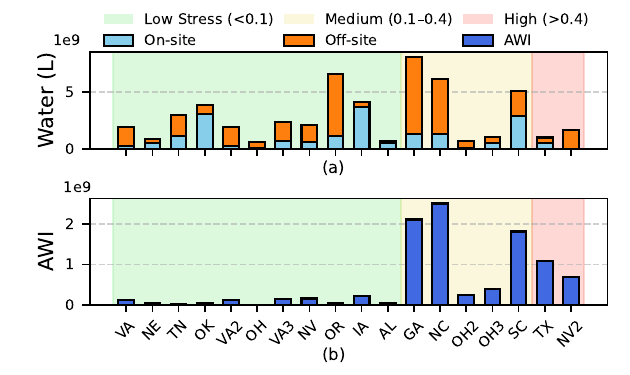}
    \caption{(a) Annual water consumption of Google's U.S. datacenters, (b) AWI of Google's U.S. datacenters using the 3\% discount rate, both grouped by regional water stress levels.}
    \label{fig:water_combined_plot}
\end{figure}

\subsubsection{Adjusted water impact in different locations.}

We begin by mapping water stress and datacenter water consumption. In \Cref{fig:us_watershed_stress_with_alaska_and_google_circles}, western areas such as Southern California and Arizona exhibit higher water stress. Many large water-consuming datacenters are located in lower-stress regions, suggesting water availability may influence site choice. Still, some datacenters are placed in high-stress watersheds, raising long-term sustainability concerns.

To dig deeper, \Cref{fig:water_combined_plot}(a) groups datacenters by their regional water stress levels and compares their annual water consumption. Following Aqueduct's classification~\cite{kuzma2023aqueduct}, we define water stress levels as: \textbf{low} (<0.1), \textbf{medium} (0.1–0.4), and \textbf{high} (>0.4). While most datacenters fall in low- or medium-stress regions, several high-stress regions still show substantial water consumption.

\SYSTEM{} then calculates the adjusted water impact by integrating both water consumption and water stress factors. As shown in \Cref{fig:water_combined_plot}(b), datacenters in medium-stress regions exhibit the highest adjusted water impact, more than 2$\times$ higher than datacenters in high-stress regions. In particular, sites in GA, NC, and SC show significantly higher adjusted water impact than those in TX or NV2, which face higher stress but consume less water.

\takeawaybox{
Water stress alone is not sufficient to capture true water impact. High water consumption in medium-stress regions can cause greater long-term impact than moderate consumption in high-stress regions.}

\subsubsection{Sensitivity check of discount rate}

\begin{figure}[!t]
    \centering
    \includegraphics[width=\linewidth]{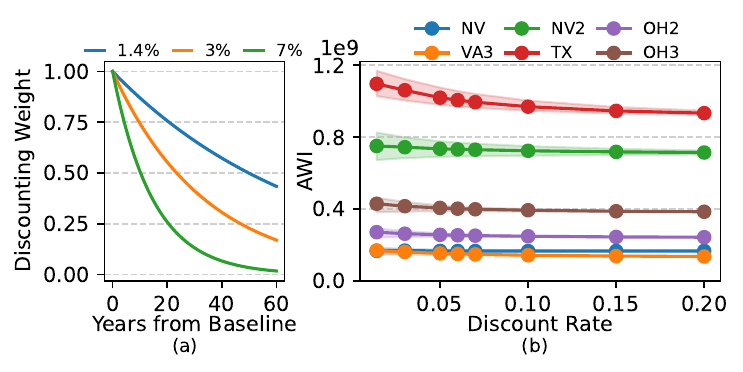}
    \caption{Sensitivity Check of Discount Rate. (a) Discounting weights of future years under different discount rates. (b) Adjusted water impact (AWI) of Google's datacenters under different discount rates, with shaded areas representing variability across different future scenarios.}
    \label{fig:combined_discount_wis_fig}
\end{figure}

The discount rate controls how much future impacts are valued compared to present ones. As shown in~\Cref{fig:combined_discount_wis_fig}(a), higher discount rates sharply reduce the weight of future years, emphasizing short-term outcomes. We adopted a 3\% discount rate in our main analysis above, following the U.S. Water Resources Development Act of 1974 (WRDA 1974)~\cite{wrda1974}.

However, discount rates vary by perspectives on intergenerational responsibility and policy planning. For example, \cite{stern2007economics} advocates a low discount rate of 1.4\% to reflect ethical responsibility to future generations. In contrast,U.S. federal budgeting often uses a 7\% rate to favor near-term returns~\cite{carter2016discount}.

\Cref{fig:combined_discount_wis_fig}(b) shows how adjusted water impact (AWI) for Google datacenters changes under different discount rates. The solid lines represent the median AWI across all future scenarios; the shaded areas reflect uncertainty among the three future pathways~\cite{kuzma2023aqueduct}. As expected, lower discount rates place more weight on long-term water stress, resulting in higher AWI and greater uncertainty with larger shaded areas. Notably, the relative comparison between datacenters can shift depending on the discount rate. For example, as the discount rate decreases, TX's AWI increases faster than NV2's, indicating that TX has higher long-term water stress. When long-term impacts are prioritized (i.e., at low discount rates), NV2 becomes the more sustainable datacenter location.


A more striking shift appears between NV and VA3. At low discount rates, NV has lower AWI, suggesting that NV is more sustainable in long term. However, as the discount rate increases, NV's AWI surpasses VA3's. This implies that when short-term water stress is more heavily weighted, VA3 is more sustainable.

\takeawaybox{
The choice of discount rate significantly alters datacenter sustainability rankings. A site that is sustainable in the long term (e.g., NV over VA3) may appear less favorable when short-term impacts are prioritized.
}



%% file: tex/5-case3.tex
\section{Case Study III: Semiconductor Fab Plants}\label{sec:case3}

In this section, we use \SYSTEM{} to evaluate the \underline{long-term} water impact of semiconductor fabrication plants (fabs), which often operate over multiple decades. For instance, several Intel fabs were built in the late 20th century and are still in active use today~\cite{tomshardware2024intel}.
 
\subsection{Evaluation Methodology}

\begin{figure}[!t]
    \centering
    \includegraphics[width=\linewidth]{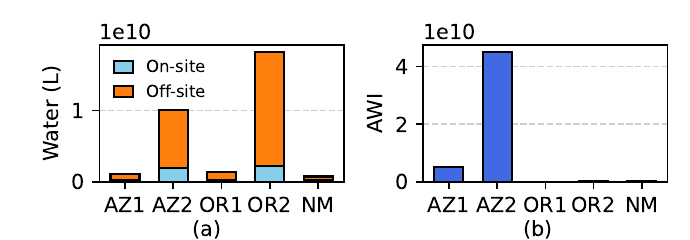}
    \caption{(a) Annual water consumption of Intel's fab plants, (b) AWI of Intel's fab plants using the 3\% discount rate.}
    \label{fig:intel}
\end{figure}

We analyze the annual water consumption of Intel's U.S. semiconductor fabs. We focus on Intel because it provides detailed public reporting, including site-specific data on annual water consumption~\cite{intelCSR2024} and energy consumption~\cite{intelExplore2024}.

\textbf{Adjusted water impact.} Given the decades-long lifespan of fab plants, \SYSTEM{} evaluates the annual adjusted water impact using long-term water stress aggregation, based on the Business-as-Usual scenario and a 3\% discount rate, as in~\Cref{sec:case2}.

\subsection{Evaluation Results}

\Cref{fig:intel} shows Intel's U.S. fabs in three main regions: Chandler in Arizona (AZ), Hillsboro in Oregon (OR), and Ronler Acres in New Mexico (NM). Both Arizona and Oregon host two campuses. \Cref{fig:intel}(a) illustrates the annual water consumption at each fab location. The OR2 site reports the highest water consumption with 1.82$\times 10^{10}$ liters per year, followed by AZ2 in Arizona. 

When \SYSTEM{} accounts for water stress, the adjusted water impact (AWI) in~\Cref{fig:intel}(b) shows significant differences. Despite consuming less water, the Arizona fabs have much higher AWI due to the region's severe water stress. In contrast, Oregon and New Mexico fabs show far lower AWI. This highlights how water stress, not just consumption volume, drives environmental impact.


\takeawaybox{
Semiconductor fabs consume a large amount of water. Intel’s Arizona fabs have higher environmental burdens than those in Oregon or New Mexico due to higher local water stress, despite lower water consumption. It emphasizes the need to factor in regional water availability in siting and operations.
}

%% file: tex/6-conclusion.tex
\section{Conclusions}
We introduce \SYSTEM{}, the first general water impact evaluation framework for computing by considering spatial and temporal variation in water stress. Using \SYSTEM{}, we assess the water impact across three layers of the computing stack: LLM serving (software), datacenters (computing infrastructure), and semiconductor fabs (hardware manufacturing). Our findings highlight the opportunities for deploying more water-sustainable computing systems.

\begin{acks}
We thank the anonymous reviewers for their valuable feedback.
This work is supported by NSF CCF-2413870.
\end{acks}